\documentclass[journal,article,submit,pdftex,moreauthors,aas_macros]{mdpi} 
\renewcommand{\linenumbers}{}
\firstpage{1} 
\pubvolume{1}
\issuenum{1}
\articlenumber{0}
\pubyear{2022}
\copyrightyear{2022}
\datereceived{} 
\dateaccepted{} 
\datepublished{} 
\hreflink{https://doi.org/} % If needed use \linebreak
\Title{Blazar jets as possible sources of ultra-high energy photons: a short review}

\TitleCitation{Title}

\Author{Gopal Bhatta $^{1,\dagger,\ddagger}$\orcidA{0000-0002-0705-6619}}

\AuthorNames{Firstname Lastname, Firstname Lastname and Firstname Lastname}

\AuthorCitation{Bhatta, G.}
\address{%
$^{1}$ \quad Institute of Nuclear Physics, Polish Academy of Sciences, PL-31342 Krak\'ow, Poland}

\corres{Correspondence: gopal.bhatta@ifj.edu.pl}

\abstract{ In this work, I present a qualitative discussion on the prospect of production of ultra-high
photons in blazars. The sources are a subclass of active galactic nuclei which host supermassive black holes and fire relativistic jets
into the intergalactic medium. The kpc scale jets are believed to be dominated by Poynting flux and constitute one of the most efficient
cosmic particle accelerators, that potentially are capable of accelerating the particles up to EeV energies. Recent IceCube detection of astrophysical neutrino emission in coincidence with the enhanced gamma-ray from Tev blazar TXS 0506 + 056 further supports hadronic models of blazar emissions in which particle acceleration processes such as relativistic shocks, magnetic re-connection, and relativistic turbulence could energize hardrons, e. g. protons, up to energies equivalent to billions of Lorentz factors. The ensuing
photo-pionic processes may then result in gamma-rays accompanied by neutrino flux. Furthermore, the fact that blazars are the dominant
source of observed TeV emission encourages us to search for signatures of acceleration scenarios that would lead to the creation of  ultra-high energy photons.}


\begin{document}

%%%%%%%%%%%%%%%%%%%%%%%%%%%%%%%%%%%%%%%%%%
%\setcounter{section}{-1} %% Remove this when starting to work on the template.

% The order of the section titles is: Introduction, Materials and Methods, Results, Discussion, Conclusions for these journals: aerospace,algorithms,antibodies,antioxidants,atmosphere,axioms,biomedicines,carbon,crystals,designs,diagnostics,environments,fermentation,fluids,forests,fractalfract,informatics,information,inventions,jfmk,jrfm,lubricants,neonatalscreening,neuroglia,particles,pharmaceutics,polymers,processes,technologies,viruses,vision

\section{Introduction}

Cosmic rays (CR) in general represent  high-energy particles that originate in the astrophysical events harboring extreme physical environments, such as supernovae, gamma-ray burst and active galactic nuclei (AGN). The constituent particles of CR mostly include charged particles  e. g. protons, fully ionized atomic nuclei and electrons, but these can include  neutral particles (e. g. neutrinos and photons) as well as neutrons at higher energies. The energy spectrum of CR can be well
 approximated by a broken power-law of the form $dN/dE \propto E^p$, where the power-law index $p$ depends upon energy; and the power-law spans over more than 13 orders of magnitude, starting from 10$^7$ eV and extending above 10$^{20}$ eV. It is an interesting fact that after more than 100 years since the discovery of cosmic rays by Victor Hess in 1912, the sources of high-energy charged cosmic rays still elude us. This is mainly due to the fact that the large scale magnetic field  intervening the galaxies can deflect the path of the CR significantly and thereby hide the source locations from us. Nevertheless, with the advancement in the detectors and computing capabilities,  we have considerably enriched our knowledge on the nature of their origin, acceleration and propagation mechanisms \cite{2019BAAS...51c..93S}. In fact, the topic is actively pursued and widely discussed in the modern high energy astrophysics.  The cosmic rays that are at the lower end of spectrum most likely have galactic origin, e. g. particle accelerated by the shocks of supernova remnants. But the ultra-high energy cosmic rays (UHECR; E$ > 10^{18}$ eV), most likely originate in extra-galactic environment. The UHECR during their propagation interact with the all-pervading cosmic microwave background (CMB) and lose a fraction of their energies such that  the CR flux is suppressed the energy $4 \times 10^{19}$ eV,  famously known as the Greisen-Zatsepin-Kuzmin (GZK) effect \cite{Greisen1966,Zatsepin1966}. This can create a horizon around $\sim 100$ Mpc from us such that super-GZK  UHECRs originating from beyond this GZK horizon are less likely to make it to the detectors on the Earth.   In general, existence of super-GZK UHECR are explained in the contexts of two classes of models: The \emph{top-down} models attempt to explain the origin of  super-GZK UHECR as a product of the decay of more energetic exotic particles, such as decay of Super Heavy Dark Matter and topological defects from the early Universe \cite{Gelmini2008,Ellis2006,Blasi2002,Birkel1998}; whereas in  \emph{bottom-up} models, particles are accelerated from lower energies up to the highest energies via various acceleration processes. Of these two models, the \emph{top-down}  models are debated as they usually predict a large of flux UHE photons which have so far not been observed by the current instruments.

UHE photons, although a small fraction of primary cosmic rays, constitute a crucial  component of UHECR. The search of UHE photons is important because while most of the UHECR, being charged particles, are deflected by the intervening  magnetic fields, UHE photons follow a direct path to the Earth leading to the identification of their sources. Moreover, creation of UHE photons is an important part of the processes involving acceleration and propagation of the charged particles, that constitute dominant fraction of primary UHECR. In other words, UHE photons are largely associated with other messenger particles like charged cosmic rays and neutrinos, and possibly gravitational waves. Therefore, the study of UHE photons supports to ongoing experimental collaboration on multi-messenger study of the Universe.

The study of UHECR in general plays a crucial role in the understanding of fundamental physical laws. It is widely considered that the Standard Model is  only represents the lower energy limit of a more fundamental theory, which ought to emerge at higher energies. The emergence of novel physics at very high energies (e. g. $>10^{14}$ GeV) is expected by Grand Unified Theories of electroweak and strong interactions. It should be emphasized that the threshold energy are so high that the current generation accelerators at the Earth  are unable to reach it. This is why the study of UHECR astrophysical origin becomes crucial to the exploration of physical laws at high energies.

In this work, I present a qualitative discussion suggesting that blazars with their relativistic jets pointed towards the Earth could be one of the  most promising sources of UHE photons.  In Section 2, some of the works focused on the search of UHE photons are highlighted. In Section 3, the possibility that blazars could be the astrophysical sources that  could produce UHE photons are discussed in detail. Finally, in Section 4, I present my conclusions.

\section{Search for UHE photons}
There are a number of works (or experiments) that are dedicated to the search of high energy photons using mainly ground based high energy telescopes. While Fermi\footnote{\url{https://fermi.gsfc.nasa.gov/}}, a space based telescope, detects gamma-ray emission up to 300 GeV regularly, MAGIC\footnote{\url{https://www.mpp.mpg.de/en/research/astroparticle-physics-and-cosmology/magic-and-cta-gamma-ray-telescopes/magic}} and HESS\footnote{\url{https://www.mpi-hd.mpg.de/hfm/HESS/}}, air shower based ground telescopes, are sensitive up to a few tens of TeV energy range. At higher energy, the Tibet Air Shower array reported detection of several gamma-ray photons with energies above 100 TeV from the Crab Nebula \cite{2019PhRvL.123e1101A}. More recently the Large High Altitude Air Shower Observatory (LHAASO) reported  detection gamma rays photon with energies about PeV energies  emitted by galactic “PeVatrons”. This detection  included a photon at 1.4 PeV,  which represents the highest energy photon event ever observed \cite{Cao2021}.
 Particles with energies more than a few tens of PeV are probably produced in large scale extra-galactic systems, such as active galactic nuclei, jets, radio galaxy lobes,  and clusters of galaxies. Super GZK UHECRs as they propagate through the Cosmic Microwave Background (CMB), interact through GZK process and produce both UHE photons, also known as GZK photons, and neutrinos via decay of neutral and charge pions, respectively. The decay of $\pi^{\pm}$ results in so-called cosmogenic neutrinos, the detection of which is one of the prime goals of some of the prominent neutrino telescopes, such as IceCube \footnote{\url{https://icecube.wisc.edu/}} and ANITA \footnote{\url{https://www.phys.hawaii.edu/~anita/}}. The decay of neutral pions ($\pi^0$) on the other hand produces GZK photons, which may carry 10\% of the energy of the UHECR that produce them, making them a significant component of UHECRs.  It is also suggested that the Universe could be more transparent to the UHE photons as compared to the UHE photons, such that photons could dominate over the other nuclei at the highest energies \citep[e. g. see][]{Risse2007,Lee1995,Akharonian1990}. However, it appears that the exact spectrum of the UHE photons would depend upon a number of factors including  the MWL extra-galactic back ground light (EBL),  the initial proton fluxes, source distributions and large-scale intervening magnetic fields.

%%The GZK process is widely believed to be one of the dominant process that produces UHE photons as well as cosmogenic neutrinos.

Generally, detection of UHE photons on the Earth relies on the methods that discriminate hadronic-induced air shower events from those initiated by UHE photons. Some of the the known properties relevant to shower evolution are utilized. The air shower initiated by a UHE photon develops more slowly compared to a hadronic cascade and therefore the maximum of the development (usually quantified by X$_{\rm max}$)  can occur close to the ground. The difference in the nature of electromagnetic cascades versus hadronic can materialize in terms of the X$_{\rm max}$ value such that it is found that on average the simulated X$_{\rm max}$ for photon induced cascades is larger by 200 g/cm$^2$ compared to cascades induced by proton \cite{Aab2017}. Moreover, the air showers initiated by UHE photons tend to show a steeper lateral distribution function, i. e. narrower in spatial extension, and show lesser muonic content. However, it should be pointed out that when the Landau-Pomeranchuk-Migdal (LPM) \citep{Migdal1956} effect, a effect which  reduces of probability of bremsstrahlung and pair creation processed at higher energies,  is included the ability to separate between UHE $\gamma$-rays and protons can be significantly weakened \citep{Wada2008}.

Although photons with energies with 1 EeV and higher have not been detected so far, 
there have been a number of efforts to estimate the upper limit of the UHE photon flux. Using the cosmic ray observations from  KASCADE and KASCADE-Grande an upper limit of the fraction of  photon with energy  $3.7\times 10^{15}$ eV to the total CR flux $1.1\times 10^{-5}$ was  estimated \citep{Apel2017}.
Similarly, EAS-MSU  \citep{Fomin2017} estimated an upper limits of the diffuse photons flux of $\sim$10 km$\rm^{-2} sr^{-1} yr^{-2}$   between the energy range of 10$^{16}$ and 3$\times 10^{17.5}$ eV.  Using the observation from the Pierre Auger Observatory (PAO) an upper limit of integral flux of UHE photons above $10^{18}$ eV  was found to be  $\sim $0.008 km$\rm^{-2} sr^{-1} yr^{-1}$ at the 95\% confidence level \cite{Aab2017}. More recently,  a search for UHE photons in the energies, i. e.  $>2\times 10^{17}$ eV was conducted using hybrid observations from the PAO. The study resulted in an upper limit of the integral photon flux  above the $10^{17}$ eV  $\rm \sim 3 km^{-2} yr^{-1} sr^{-1}$ \cite{Abreu2022}.

%Only three photon candidates at energies 1–2 EeV are found, which is compatible with the expected hadroninduced background. From \cite{Aab2017}

Study of particle cascades initiated by UHE photons in the geomagnetic field has been studied using  Monte Carlo simulations \cite{Homola2013,Homola2005}. Similarly, propagation of UHE photons in the solar magnetosphere was also studied using Monte Carlo simulations by \cite{Dhital2022,2020APh...12302489A,Alvarez-Castillo2022,Poncyljusz2022}. The results showed that  the photon disintegrates giving rise to an extended cascades comprising of thousands of spatially correlated secondary particles which lose energy through synchrotron emission. On the Earth the observational signature of such extended cascade can be searched in the form of temporal clustering of the cosmic ray events \cite{Clay2022}. In addition, a global collaboration named The Cosmic Ray Extremely Distributed Observatory (CREDO) \cite{Homola2020} is  dedicated to searching the footprints of such spatially correlated cosmic ray events that might have initiated by UHE photons.

\section{Blazar jets possible source of UHE photons}
AGN powered by supermassive black holes represent some of the largest energy reservoir in the Universe. Energy extracted from the black hole can launch powerful relativistic jets, which stream particles with the speeds comparable to the speed of light. Blazars are a subclass of AGN that feature relativistic jets aligned close to the line of sight \cite{1995PASP..107..803U}. As a result, the emission is significantly Doppler boosted and rapidly variable. These sources  are characterized with high luminosity, broadband emission and  variability in all temporal and spatial frequencies. The broadband emission  from blazars can be detected across the entire observable electromagnetic spectrum, from radio to TeV $\gamma$-rays. Blazars can be further classified into flat-spectrum radio quasars (FSRQs)
and BL Lac objects based upon the presence or absence of emission lines, respectively. The broad-band spectral energy distributions can be recognized by two  hump-like non-thermal emission components \cite{1998MNRAS.299..433F}. The low-frequency component lying between radio and X-rays is widely accepted as being due to synchrotron emission by relativistic electrons in the jet; and the
high-frequency  component peaking between X-rays to $\gamma$-rays can be ascribed to inverse
Compton scattering of low-energy target photons \cite{Blandford2019}. The sources also are sub-divided  as low-synchrotron-peaked, intermediate-synchrotron-peaked and high synchrotron-peaked   based on the location of the peak of the synchrotron component ($\nu_{sy}$) , that is, $\nu_{sy}$, that is, $\nu_{sy} < 10^{14}$ Hz,  $10^{14} \  Hz < \nu_{sy}  < 10^{15}$ Hz and  $\nu_{sy}  > 10^{15}$ Hz, respectively \cite{2010ApJ...722..520A}.

Blazars jets fueled by  $\sim 10^9$ M$\rm _{\odot}$  black holes  are widely considered as candidate sources that are capable of accelerating  particles up to UHE range \citep{Rodrigues2021,Tursunov2020}. It is highly likely that the CRs with energies about 10$^{20}$ eV, nearly seven orders of magnitude more than the energy of the particles produced at the Earth accelerators, probably are originated  from the sources that belong to AGN family.  Radio-loud AGN with their large scale (kpc/Mpc) jets, satisfying the Hillas criterion $E_{max}=qBR$ that would be required to fulfill by potential acceleration sites, provide most promising venues that is favorable to the particle acceleration up to the EeV energies.  In fact, blazars are the dominant discrete sources that contribute the  TeV gamma-ray emission observed from the Earth. Moreover,  the high-energy neutrino event IC 170922A was found to coincide with an enhanced $\gamma$-ray emission from the TeV  blazar TXS 0506+056 \cite{2018Sci...361.1378I,2018Sci...361..147I}. The coincidence bolstered that idea that blazars host the condition favorable to accelerate the particles that is required to produce PeV neutrino and therefore likely to be most probable source class associated with discrete neutrino events. Although the exact production mechanism of neutrino in blazars is still debated, the particles are closely linked to the production UHECRs. Several possible scenarios leading to the neutrino emission from  blazar have been extensively discussed in several works \cite[e. g. see][]{Bednarek1999,Atoyan2001,Dermer2014,Petropoulou2016,Kreter2020}. Highly accelerated protons on interaction with the ambient medium to give rise to $pp$ interaction leading to  the neutrino production \citep{Schuster2001}. Also, interaction of high energy protons with the internal and external photon fields might results in photo-production of pion, which subsequently decay to neutrino emission\citep{Mannheim1995,Dermer2009}

Following bottom-up scenario, UHE photons can be produced in blazar jets in the course of particle acceleration of CRs up to very high energies. Therefore, they are most likely to originate near the regions that are co-spatial to the population of UHE charged particle. The UHE protons (or in general charged UHECR) lose energy via mainly proton-proton (pp; $p+p\rightarrow  \pi^{0/\pm}+X$) and proton-photon ($p\gamma$) collisions. These interactions result in charged or neutral pions which further decay  producing neutrinos and gamma-rays, respectively. Furthermore, for  a $\gtrsim E \times10^{19.5}$ eV proton on interaction with the photon background generates
a pion, which rapidly decays into two photons or an electron/muon and  
neutrino depending on its charge ($p+\gamma \rightarrow p/n + \pi^{0/+}$); whereas  UHE proton with energies $\lesssim E \times10^{19.5}$ eV,  the photo-pion interaction yields electron-positron  pair ($p+\gamma \rightarrow p+e^{+}+e^-$) which consequently   loses energy via synchrotron emission. In addition, UHECR can induce cascade radiation in the extragalactic background light resulting in a high energy photon field with energies as high as 10 EeV. The signatures of such processes can be observed in the spectra of luminous blazars  \cite{2009NJPh...11f5016D}. Also photomeson interaction of neutrons outside the blob can produce  UHE photons which, along with neutrons and neutrinos, can  escape the BLR to form a well collimated neutral beam of the jet. It is also suggested that protons can be accelerated to UHE range close to the inner jet region such that these UHECR in turn power a neutral beam of neutrinos, neutrons, and $\gamma$-rays from p-$\gamma$ photopion production. \cite{Dermer2012}. In fact,  the transportation of energy  via beams of energetic neutral particles, e. g. neutrons and photon beams,   could be one of the feasible means to power the hot spots and lobes visible in radio and X-ray frequencies \citep{Bednarek1999}. However, in some cases the  production an excess of neutrinos via photo-pionic processes may require a large target photon densities making the source opaque to high-energy gamma-rays \citep{Halzen2019}.  This might raises  some doubt about the idea that the two different messengers could originate from the same location.
 
The production of UHE photons are linked to the particle acceleration mechanism that can accelerate the charge particles up to EeV range; and of the several particle acceleration mechanisms possible in the relativist jets of radio-loud AGN, some of the widely discussed scenarios are based on relativistic shocks, magnetic re-connection and turbulence. In the weakly magnetized jets, relativistic shocks are dominant mechanism to accelerate the particles to very high energies \citep{Kirk1998}.  At the shock front, following Fermi acceleration, the particles gain an energy average energy of  $\left \langle \Delta E/E \right \rangle=4/3\left ( v/c \right )$ per crossing,  as they frequently cross the shock wave front back and forth.  In the finite shock scenario where the maximum energy attained by a proton in a finite shocks with speed $\beta_s$ can be given as   
\begin{equation*}
 E_{p,max} \simeq 7.8\times 10^{20} \beta_s \left ( B/G \right )\left (r_\perp /pc\right ) eV, 
 \end{equation*}
  it is easy to see that as the shock travel a pc scale distance in the blazar jet with a typical magnetic field 1 G, the proton is accelerated up to energies of tens of EeV \cite[see][]{Mannheim1993} .

However, it may not be possible for a single shot of relativistic shocks in the radio jets to accelerate even the lighter charged particle to UHE domain \citep{Bell2018}. Therefore, it is suggested that lower-energy CRs, such as the ones in produced in supernova remnants may enter the AGN jet and boosted to UHE domain through multiple shock events,  yielding UHECR emission that may be beamed along the jet axis \citep{Mbarek2019}.  An in-depth analysis exploring the nature of the variability of gamma-ray emission from a sample of blazars using decade-long  Fermi/LAT (100 MeV--300 GeV) observations showed that the power spectrum density (PDS)  slope indexes  of the variable gamma-ray emission were found to be $\sim$ 1, and the flux histograms were best described lognormal probability distribution function (PDF) \cite{2020ApJ...891..120B}.  The PSD with slope index $\sim$ 1, famous as in flicker noise, could imply that the variability is generated by a long-memory process; and the lognormal PDF is often interpreted as multiplicative process. Such a result provides evidence that particles emitting gamma-rays retain imprints over many orders of temporal and fluxes scales. Together this supports the idea that it might take the entire spatial extent of the jets for the particles to gain very high energies.

In the case of highly magnetized jets, the conditions are favorable for the magnetic re-connection events \cite{2022arXiv220409100D}. Such events take place  when two oppositely polarized magnetic region in conduction plasma collide in the jets;  consequently energy is released as magnetic lines partially break and rearrange for stable configuration. Magnetic re-connection events  rampant along the jets can accelerate the particle up to very high energies. Compared to a gradual acceleration at the shock wave front the magnetic re-connection are sudden and  rapid in nature \cite{2006Natur.443..553D}. If fact, the minute timescale variability in the TeV emission  as observed by the HESS supports such a fast outburst emission from extremely compact regions moving with Lorentz factors that are much larger than the typical bulk Lorentz factors 10--20  \cite{Aharonian2007}. Magnetic re-connection events might  in some cases also lead to the formation of mini-jets within the main bulk jet. This has the net effect of boosting the energy of the particles, including photons,  by Doppler factors according to  $\Gamma =\Gamma _1\Gamma _2(1+v_1v_2cos\theta )$, where $\Gamma$ is the resultant Lorentz factor of the particles in the mini jet that is ejected with $\Gamma _2$ along an angle $\theta$ with the main jet propagating with $\Gamma _1$  bulk Lorentz factor \cite{Giannios2009}. 

In blazar jets  a population of photons might also undergo multiple inverse-Compton scattering by co-spatial UHECRs, thus gradually raising the energy of the photons to  very high energies. A UHE photon later (after some optical length) might disintegrate in the locally strong magnetic field, by pre-shower effect, resulting in the extended shower of particles. This population of secondary particles with lower energy can lose energy by synchroton emission creating blazar flares visible over multiple wavebands. Therefore, the flares that appear (quasi-)simultaneously in several wave bands could be an evidence that they are likely to result from the disintegration of a ensemble of high energy photons. A strong correlation observed in  \cite{Bhatta2021} might be an observational signature of the such processes. Similarly, in a turbulent jet scenario particles can be stochastically accelerated along the jet \citep[e. g. see][]{Mbarek2021} as magnetized cells move randomly, mimicking the magnetic mirrors as in the second order Fermi acceleration scenario \cite{2014ApJ...780...64A,2014ApJ...780...87M}.

%%%%%%%%%%%%%%%%%%%%%%%%%%%%%%%%%%%%%%%%%%
\section{Conclusions}
The search for UHE photons of astrophysical origin is one of the actively pursued research in field of high energy astrophysics. Also, the search for UHE photons is a part of the multi-messenger approach to the study of violent astrophysical events in the Universe. These studies provide insights into some of the most exciting topics including neutron stars, binary black holes, AGN jets and large scale structure such as galaxy clusters. UHE photons mainly result from the decay of the neutral pions originating in hadronic interactions involving UHECR.  In such a context, it is natural to consider blazars as the most promising candidate source class to produce UHE photons. The kpc scale relativistic jets of blazars host  some of the most violent particle acceleration processes, such as relativistic shocks, magnetic reconnection and turbulence,  making them most efficient cosmic  particle accelerators. Long memory processes, log-normal flux distribution of gamma-ray flux, and correlation among MWL emission from blazars can potentially provide  observational signatures of the conditions in which UHE photons can originate. More quantitative treatment on the subject follow in the future works.

%%%%%%%%%%%%%%%%%%%%%%%%%%%%%%%%%%%%%%%%%%
\vspace{6pt} 

%%%%%%%%%%%%%%%%%%%%%%%%%%%%%%%%%%%%%%%%%%
%% optional
%\supplementary{The following supporting information can be downloaded at:  \linksupplementary{s1}, Figure S1: title; Table S1: title; Video S1: title.}


%%%%%%%%%%%%%%%%%%%%%%%%%%%%%%%%%%%%%%%%%%

\funding{I acknowledge financial support by the Narodowe Centrum Nauki (NCN) grant UMO-2017/26/D/ST9/01178.}

\acknowledgments{I would like to thank the anonymous referees for their constructive comments
and suggestions that greatly improved the manuscript.}

\dataavailability{Not applicable}

\conflictsofinterest{The author declares no conflict of interest.}

%% Only for journal Encyclopedia
%\entrylink{The Link to this entry published on the encyclopedia platform.}

\abbreviations{Abbreviations}{
The following abbreviations are used in this manuscript:\\

\noindent 
\begin{tabular}{@{}ll}
AGN & Active galactic nuclei\\
BLR & Broad-line region\\
CR& Cosmic rays\\
CREDO & The Cosmic Ray Extremely Distributed Observatory \\
EBL & Extra-galactic background light\\
HESS &High Energy Stereoscopic System\\
LPM & Landau-Pomeranchuk-Migdal \\
MAGIC &  Major Atmospheric Gamma Imaging Cherenkov Telescope\\
MC & Monte Carlo \\
PDF & Probability Density Function \\
PSD & Power Spectral Density \\
UHE & Ultra-high energy \\
UHECR & Ultra-high energy cosmic rays
\end{tabular}
}

%%%%%%%%%%%%%%%%%%%%%%%%%%%%%%%%%%%%%%%%%%
\begin{adjustwidth}{-\extralength}{0cm}

\reftitle{References}

\end{adjustwidth}
\end{document}